# Spin glass state and enhanced spiral phase in doped delafossite oxide $CuCrO_2$

Z. R. Yan[1], M. H. Qin[1] [*], S. Dong[2], M. Zeng[1], X. B. Lu[1], X. S. Gao[1], and J. –M. Liu[3] [*]

[1]*Institute for Advanced Materials and Guangdong Provincial Key Laboratory of Quantum Engineering and Quantum Materials, South China Normal University, Guangzhou 510006, China*

[2]*Department of Physics, Southeast University, Nanjing 211189, China*

[3]*Laboratory of Solid State Microstructures, Nanjing University, Nanjing 210093, China*

**[Abstract]** In this work, we study the doping effects on the magnetic states of $CuCrO_2$ based on the classical frustrated spin model [Lin *et al*., Phys. Rev. B 89, 220405(R) (2014)]. Several experimental observations can be well reproduced by the Monte Carlo simulations of the modified spin models. Our work suggests that the disorder induced by V/Al doping cooperated with the frustration in the system may contribute to the emergence of the spin glass state. Furthermore, the hole-doping by $Mg^{2+}$ substituting $Cr^{3+}$ enhances the quantum fluctuations and bond disorder which modulate the biquadratic exchanges, and in turn results in the promotion of the spiral phase, consistent with the experimental report.

[*] Electronic mail: qinmh@scnu.edu.cn, and liujm@nju.edu.cn

## I. Introduction

During the past decades, nontrivial magnetic phases observed in so-called type-II multiferroic materials such as the delafossite oxide $CuCrO_2$ have drawn extensive attentions due to their interesting physics and potential applications.[1-6] Specifically, the magnetic $Cr^{3+}$ ions (having quasiclassical $S = 3/2$ spin) in $CuCrO_2$ form a triangular lattice in the *ab* plane, and the compound exhibits an incommensurate proper-screw spiral state (ICY state) below the critical temperature $T_N$. In this state, the spiral plane is perpendicular to the *ab* layer, and the three spins in each triangle sublattice form the so-called 120° structure, as shown in Fig. 1(a). Interestingly, an electric polarization (*P*) along the spiral propagation vector *k* is induced through the variation of the hybridization between the Cr *d* orbitals and the O *p* orbitals caused by spin-orbit coupling.[6,7]

Owing to the strong magnetoelectric coupling in this system, a significant dependence of *P* on the magnetic field **H** has been observed in experiments.[8-10] For example, the flop of *P* from the *x*-axis to the *y*-axis has been reported due to the 90° rotation of the spin-spiral plane under **H** applied along the *y* direction.[11] Theoretically, several **H**-induced cycloidal spiral phases have been uncovered in the anisotropic classical spin model for $CuCrO_2$,[12,13] well consistent with the experimentally reported electric polarization based on the Arima's mechanism for multiferroic behavior.[7] The competitions between magnetic frustration, Zeeman energy, and thermal fluctuations are suggested to be responsible for the changes of magnetoelectric properties in $CuCrO_2$ under **H**.[13]

On the other hand, the magnetic states of $CuCrO_2$ can also be effectively modulated through impurity doping, and several interesting phenomena have been reported in experiments.[14] For example, a spin glass state induced by the $V^{3+}$ for $Cr^{3+}$ substitution has been observed in $CuCr_{1-x}V_xO_2$ series for $x > 0.18$.[15,16] Similarly, a short-range antiferromagnetic (AFM) excitation resulted from the enhancement of spin glass component has been observed in nonmagnetic Al-doped systems.[17] More interestingly, significant hole-doping effects have been uncovered in $CuCr_{1-x}Mg_xO_2$ for $x \leq 0.03$. In detail, $T_N$ shifts toward high *T* with the substitution of nonmagnetic $Mg^{2+}$ for $Cr^{3+}$, indicating the important role of the coupling between the itinerant hole and localized spin.[18-21]

The study on the doping effects on magnetic states in multiferroic oxides becomes very important from the following two viewpoints. On one hand, this study helps one to understand the multiferroic physics and to search for more attractive systems with improved magnetoelectric performance. On the other hand, despite the long history of research, spin glass phase transition is still a hot topic in statistical mechanics, while few results have been reported in triangular antiferromagnets.[22] Thus, the study of doped $CuCrO_2$ is essential both in application potential and in basic physical research. However, several doping effects on magnetic phase transitions reported in experiments are still far from well understood in theory, including: 1) spin glass behaviors reported in $CuCr_{1-x}V_xO_2$ and $CuCr_{1-x}Al_xO_2$, and 2) the unconventional promotion of the ICY state by the hole-doping for the Mg substitution.

Fortunately, the earlier spin model for $CuCrO_2$ which has been successfully used to explain the magnetic field effects allows one to explore the doping effects based on a modified model.[13] For example, in our earlier work, both the lattice defects and random exchange induced by the isovalent substitution of nonmagnetic $Al^{3+}$ for $Cr^{3+}$ are confirmed to be responsible for the decrease of $T_N$ with the increasing Al-doping magnitude.[23,24] In this work, we study the modified spin models to further investigate the doping effects on the magnetic phase transitions in doped $CuCrO_2$. Several experimental observations are well explained in our Monte Carlo (MC) simulations. In detail, it is suggested that 1) the spin glass state in $CuCr_{1-x}V_xO_2$ is resulted from the competitions between the AFM $Cr^{3+}$-$Cr^{3+}$ coupling and ferromagnetic (FM) $V^{3+}$-$Cr^{3+}$ coupling, and 2) the hole-doping by $Mg^{2+}$ for $Cr^{3+}$ enhances the quantum fluctuations and bond disorder, and modulates the biquadratic interactions, and in turn enhances the ICY state.

The remainder of this paper is organized as follows: in Sec. II the model and the simulation method will be described. Sec. III is attributed to the simulation results and discussion, and the conclusion is presented in Sec. IV.

## II. Model and method

For the case of $V^{3+}$ substituting for $Cr^{3+}$, no additional hole is produced in the system, while the competing interactions between $Cr^{3+}$ and $V^{3+}$ are available (strong AFM $Cr^{3+}$-$Cr^{3+}$ coupling, weak AFM $V^{3+}$-$V^{3+}$ coupling, and FM $Cr^{3+}$-$V^{3+}$ coupling, as depicted in Fig. 1(b)).[16] Taking into account this fact, we studied a modified spin model for $V^{3+}$ randomly

doped CuCrO$_2$, and the Hamiltonian can be written as:

$$H = \sum_{\langle i,j \rangle} J_{ij} \cdot \Delta_{ij} \cdot S_i \cdot S_j + \sum_i \frac{1}{2} A_x S_{i,x}^2 - \sum_i \frac{1}{2} A_z S_{i,z}^2 - \sum_i \mathbf{H} \cdot S_i, \tag{1}$$

with

$$\Delta_{ij} = \begin{cases} 1 & Cr^{3+} - Cr^{3+} \\ -0.4 & V^{3+} - Cr^{3+} \\ 0.2 & V^{3+} - V^{3+} \end{cases}, \tag{2}$$

The first term is the exchange interaction between the nearest neighbors, and a spatial anisotropy with $J'/J = 0.7654$ is considered, as shown in Fig. 1(a). The couplings between different ions are modulated by the parameter $\Delta_{ij}$ which are chosen to be consistent with the experimental report. The second term is the in-plane hard-axis anisotropy with $A_x = 0.005J$, and the third term is the out-of-plane easy-axis anisotropy with $A_z = 0.05J$, and the last term is the Zeeman coupling. Here, $J'/J$, $A_z$ and $A_x$ are chosen to be the same as those in earlier work,[13] which well reproduce the spin state under zero $\mathbf{H}$. For simplicity, the length of $S$ ($Cr^{3+}$ and $V^{3+}$), $J$ and the Boltzmann constant are set to unity.

Several parameters are calculated in order to characterize the ICY state and spin glass state. For example, the vector chirality is calculated by:[25]

$$\chi = \frac{2}{3\sqrt{3}} \frac{1}{N} \sum_{\mathbf{r}} (S_A \times S_B + S_B \times S_C + S_C \times S_A), \tag{3}$$

where $N = L \times L$ is the amount of the total spins, and the sum is over all the plaquettes of the system (ABC in Fig. 1). The components parallel ($\chi_\parallel$) and perpendicular ($\chi_\perp$) to $\mathbf{H}$ or to the easy $z$ axis ($\mathbf{H} = 0$) are calculated. For the ICY state, $\chi_\perp > 0$ and $\chi_\parallel = 0$ are expected. For the spin glass transition, the order parameter generalized to wave vector $\mathbf{k}$ is defined to be: $q^{\mu\nu}(\mathbf{k}) = N^{-1}\Sigma_i S_i^{\mu(1)} S_i^{\nu(2)} \exp(i\mathbf{k}\mathbf{R}_i)$, where $\mu$ and $\nu$ are spin components, and "(1)" and "(2)" denote two identical copies of the system with the same interactions, and $\mathbf{R}_i$ is the position vector at $i$ site.[26,27] Then, the spin glass susceptibility is calculated by: $X_{SG}(\mathbf{k}) = N\Sigma_{\mu,\nu} [\langle |q^{\mu\nu}(\mathbf{k})|^2 \rangle]_{avg}$, where $\langle\ldots\rangle$ denotes the thermal average and $[\ldots]_{avg}$ is the average over disorder. In this work, disorder averages are taken over $N_s$ samples, with $N_s$ ranging from 400 for $L = 6$ to $N_s = 10$ for $L = 24$. Subsequently, the spin glass correlation length is determined from:

$$\xi_L = \frac{1}{2\sin(k_{min}/2)}\left(\frac{X_{SG}(0)}{X_{SG}(\mathbf{k}_{min})}-1\right)^{\frac{1}{2}}, \qquad (4)$$

where $\mathbf{k}_{min} = (2\pi/L)(1, 0, 0)$. Thus, the freezing temperature $T_g$ is estimated from the crossing points of $\xi_L/L$ for different $L$, according to the scaling law $\xi_L/L = f(L^{1/\nu}(T - T_g))$, where $\nu$ is the correlation length exponent.

On the other hand, itinerant holes are produced by $Mg^{2+}$ for $Cr^{3+}$ substitutions, enhancing the quantum fluctuations and bond disorder. In earlier works, it has been clearly proved that thermal fluctuations and bond disorder can produce an effective biquadratic exchange in the classical Heisenberg triangular antiferromagnet.[28] Furthermore, quantum fluctuations can also generate a similar term, as uncovered by the perturbation theory in earlier work.[29] Thus, the biquadratic interactions are further taken into account in the model to study the $Mg^{2+}$ doping effects:

$$H_K = K_1\sum_{\langle i,j\rangle}(S_i\cdot S_j)^2 + K_2\sum_{[i,k]}(S_i\cdot S_k)^2. \qquad (5)$$

Here, the nearest neighbor and the next nearest neighbor biquadratic interactions are considered. Furthermore, $Cr^{4+}$ cations are also produced due to the additional holes, and a double-exchange FM interaction between $Cr^{4+}$-$Cr^{3+}$ may be available.[21] In this work, we set $\Delta_{ij} = -1$ for $Cr^{4+}$-$Cr^{3+}$ interaction, and neglect the lattice defects and the itinerancy of $Cr^{4+}$ due to the very small amount of $Mg^{2+}$.

Our simulation is performed using the standard Metropolis algorithm and temperature exchange method.[30,31] Unless stated elsewhere, the simulation is performed on a $24 \times 24$ lattice with periodic boundary conditions.

### III. Simulation results and discussion

*A. Spin glass state induced by magnetic/nonmagnetic impurity doping*

First, we study the effects of the V-doping on the magnetic properties in CuCrO$_2$. Fig. 2(a) shows the simulated $\chi_\perp$ as a function of $T$ for various $x$. In the clean limit $x = 0$, when $T$ falls down to the transition point, $\chi_\perp$ increases, while $\chi_\parallel$ and the spin coplanarity remain small (not shown here), figuring the development of the ICY state. With the increase of $x$, the $\chi_\perp$ curve shifts toward low $T$ side, demonstrating the suppression of the ICY state. The transition point $T_N$ ($T_{peak}$, exactly) can be roughly estimated from the position of the peak in the calculated specific heat $C$, as given in Fig. 2(b). It is clearly shown that $T_{peak}$ decreases with the increasing $x$, consistent with the experimental report.[16] Furthermore, the value of $\chi_\perp$ is also significantly decreased, indicating that the ICY state is not dominated at low $T$ (for $x > 0.2$, at least).

Interestingly, spin glass order at low $T$ emerges due to the combination of disorder and frustration for $x > 0.2$. For example, Fig. 3(a) shows the calculated $\xi_L/L$ as a function of $T$ for various $L$ at $x = 0.5$. From the common well-defined crossing point, we estimate the freezing temperature $T_g = 0.087 \pm 0.005$. Actually, the finite temperature spin glass transition has been reported in a dilute Ising system on the triangular lattice.[22] In this work, it is suggested that the V doping produces disorder, and leads to spin glass behavior in the frustrated Heisenberg model with the uniaxial anisotropy.[32] Furthermore, we plot $\xi_L/L$ in the scaling form in Fig. 3(b), and estimate that the spin glass transition is with a critical exponent $v = 1.25 \pm 0.03$.

As a short summary, the simulated phase diagram for V doping is presented in Fig. 4(a) which qualitatively reproduces the experimental one. The conventional ICY state is suppressed with the increase of $x$, and a spin glass state is favored beyond $x > 0.2$ due to the introduction of disorder. Furthermore, the spatial anisotropy may be changed with $x$ in real materials. Thus, the case of the spatial isotropic model ($J' = J$) is also investigated, and the corresponding results are shown in Fig. 4(b). It is noted that the frustration is further enhanced in the model for $J' = J$, in favor of spin glass magnetism, resulting in the enlargement of the spin glass phase in the phase diagram. However, the estimated $v$ for $x < 0.2$ are rather abnormal ($v > 2$), indicating that the spin glass behavior is a little different from that for $x > 0.2$. Furthermore, the calculated $\chi_\perp$ curves show that the ICY state can be well stabilized at low $T$ for $x < 0.2$ (the corresponding results are not shown here), further demonstrating that the estimated $T_g$ in this $x$ region is not a genuine one and additional disorder is needed for the development of the spin glass state.

On the other hand, it has been experimentally reported that the spin defects produced by Al doping destabilizes the ICY state accompanying the enhancement of the spin glass component.[17] This behavior is also reproduced in our simulations in which the nonmagnetic $Al^{3+}$ impurity is simply considered as a lattice defect. The spin-glass order can be observed at low $T$ when $x$ increases above 0.35, as shown in Fig. 5(a). Similarly, the case of $J' = J$ is also investigated, and a rough phase diagram in ($x$, $T$) parameter plane is presented in Fig. 5(b). The spin glass state shows significantly dependence on the spatially anisotropy, which may provide useful information in understanding the experimental observations.

In short, our work in this part undoubtedly demonstrates the important role of disorder caused by V/Al doping in the development of the spin glass state in $CuCrO_2$, although the accurate determination of transitions points may be not available due to the finite-size effects in the simulations.

### B. Enhanced ICY state in Mg-doped system

At first glance, the introductions of the lattice defects and FM $Cr^{4+}$-$Cr^{3+}$ interaction by the substitution of nonmagnetic $Mg^{2+}$ for $Cr^{3+}$ will definitely destabilize the ICY state, contrary to the experimental observation. Thus, the nontrivial promotion of the ICY state in Mg doped $CuCrO_2$ indicates the essential role of the doped holes in modulating the magnetic phase transition. It is expected that the interaction between the doped hole and the localized spin may enhance the quantum fluctuations which could be described by the classical biquadratic interactions (shown in Equation (5)).[25] Here, we introduce the additional biquadratic interactions in the model to study the hole-doping effects. Furthermore, in one of the earlier theoretical works studying a similar triangular antiferromagnet, it has been proved that quantum fluctuations may produce an effective biquadratic exchange (negative $K_1$ and $K_2$), while bond disorder may generate a positive $K_1$ interaction.[28] However, the exact values of $K_1$ and $K_2$ are not available so far, and we systematically studied the effects of $K_1$ and $K_2$ on the multiferroic phase transition in this work.

Fig. 6(a) shows the calculated specific heat curves for various ($K_1$, $K_2$). As the magnitude of the negative $K_2$ increases from zero ($-K_2 < 0.08$), the transition to the ICY state shifts toward high $T$ side. It is easily noted that the ICY state can be further stabilized by the negative $K_2$ interactions due to the fact that the angle between the next nearest neighbors is much less than $\pi/2$. The simulated results are summarized in Fig. 6(b) which presents the detailed phase diagram in the ($K_1$, $K_2$) plane. Considering the small amount of Mg-doping in experiments, the effective biquadratic exchanges are expected to be much weaker than the spin exchanges. Thus, it is strongly suggested that the cooperation of the enhanced quantum fluctuations and bond disorder may cover the negative effect of the FM $Cr^{4+}$-$Cr^{3+}$ interaction and enhance the ICY state, leading to the increase of $T_{peak}$, qualitatively consistent with experimental observation.[19]

Furthermore, earlier experiments reveal that $T_{peak}$ for $x = 0.03$ shifts toward low $T$ by applying a 9 T magnetic field, while that for $x = 0$ is almost unaffected, demonstrating an improved magnetoelectric properties in Mg doped system.[19] This interesting phenomenon is also captured in our simulations, and the calculated results are shown in Fig. 7 which well reproduces the experimentally observations. In the clean limit $x = 0$, the specific heat curves for $H_y = 0$ and $H_y = 0.3$ (magnetic field applied along the $y$-axis) coincide with each other (Fig. 7(a)), indicating that $H_y = 0.3$ never affect the AFM transition behavior. For $x = 0.03$, $T_N$ shifts toward low $T$ side and the peak height of $C$ is significantly decreased by applying $H_y = 0.3$. Thus, it is strongly suggested that the FM $Cr^{4+}$-$Cr^{3+}$ interaction and biquadratic exchanges induced by $Mg^{2+}$ doping are responsible for the magnetic field effects on $T_{peak}$ in the hole-doped $CuCrO_2$.

## IV. Conclusion

In conclusion, we have studied the doping effects on the magnetic states of $CuCrO_2$ by the Monte Carlo simulation of the frustrated spin models. It is suggested that the disorder induced by V/Al doping and the frustration in the system may result in the emergence of the spin glass state, consistent with the experimental observations. Furthermore, the hole-doping by $Mg^{2+}$ substituting $Cr^{3+}$ enhances the quantum fluctuations and bond disorder, and in turn leads to the unconventional promotion of the AFM ICY state.

**Acknowledgements**:

The work is supported by the Natural Science Foundation of China (Grants No. 51332007, No. 51322206, No. 11274094), and the National Key Projects for Basic Research of China (Grant No. 2015CB921202), and the National Key Research Programme of China (Grant No. 2016YFA0300101), and the Science and Technology Planning Project of Guangdong Province (Grant No. 2015B090927006).

**FIGURE CAPTIONS**

Figure 1. (color online) Spin structure and exchange interaction in clean system (a) and doped system (b). The blue/red dotted lines denote the AFM/FM exchanges, and red circle is $V^{3+}/Cr^{4+}$ cation. The three sublattices are labeled by $A$, $B$, and $C$, and the wave vector $k$ and induced polarization $P$ are also shown with the red arrow.

Figure 2. (color online) The calculated $\chi$ (a) and specific heat $C$ (b) as a function of $T$ for various $x$.

Figure 3. (color online) The calculated $\xi_L/L$ (a) as a function of $T$, and (b) as a function of $L^{1/\nu}(T-T_g)$ for various $L$ at $x = 0.5$.

Figure 4. (color online) The estimated phase diagram in the ($x$, $T$) plane for V doping with spatial anisotropy (a) and isotropy (b). The correlation exponents are also depicted.

Figure 5. (color online) The estimated phase diagram in the ($x$, $T$) plane for Al doping with spatial anisotropy (a) and isotropy (b). The correlation exponents are also depicted.

Figure 6. (color online) The calculated $C$ as a function of $T$ for various ($K_1$, $K_2$) (a) and the contour plot of $T_{peak}$ in the ($K_1$, $K_2$) parameter plane (b).

Figure 7. (color online) The calculated specific heat $C$ as a function of $T$ for $x = 0$ at $H_y = 0$ and $H_y = 0.3$ ($K_1 = 0$, $K_2 = 0$), and (a) $x = 0.03$ at $H_y = 0$ and $H_y = 0.3$ ($K_1 = 0.05$, $K_2 = -0.02$) (b).

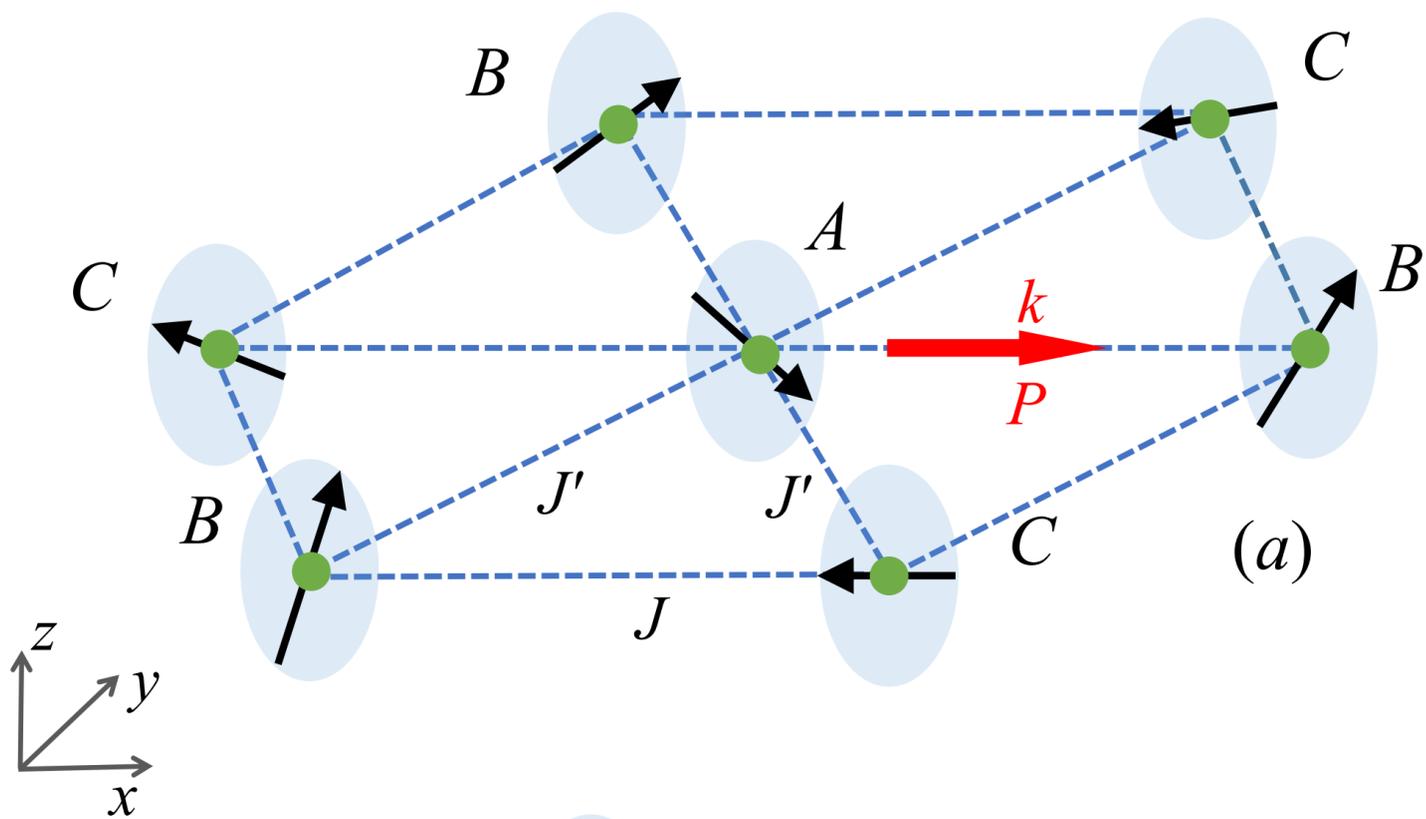

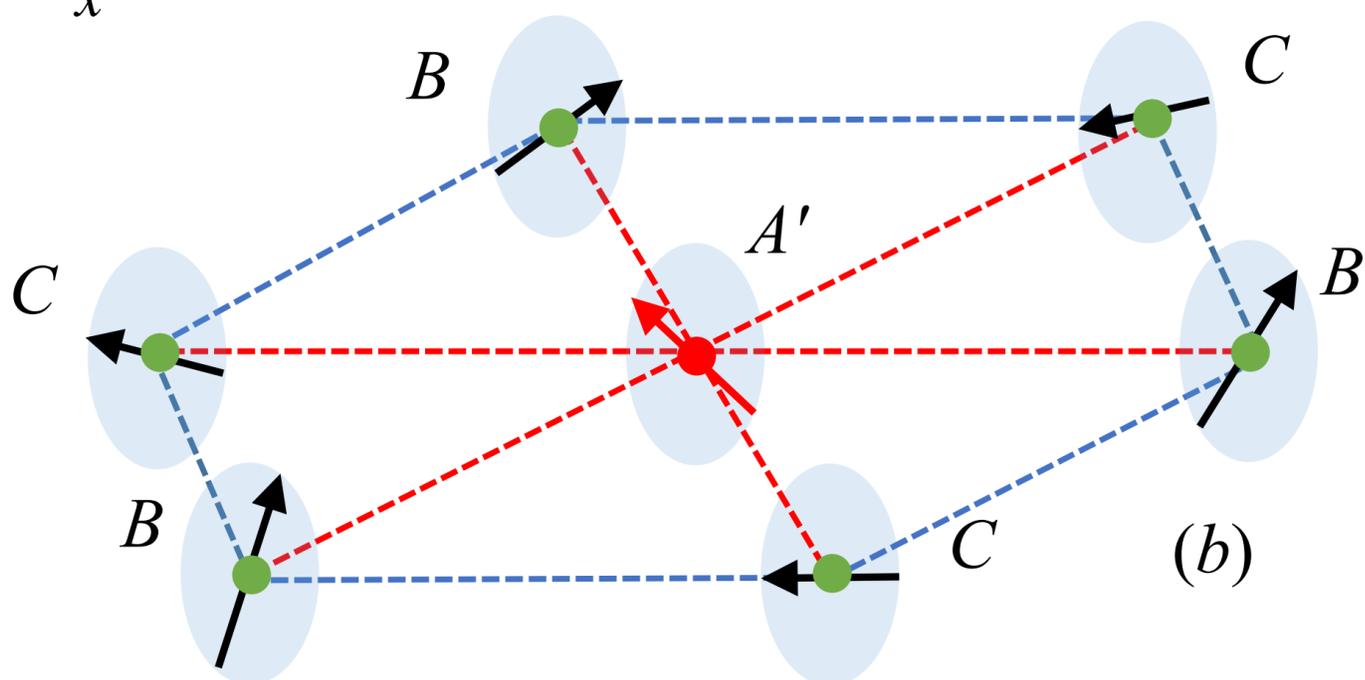

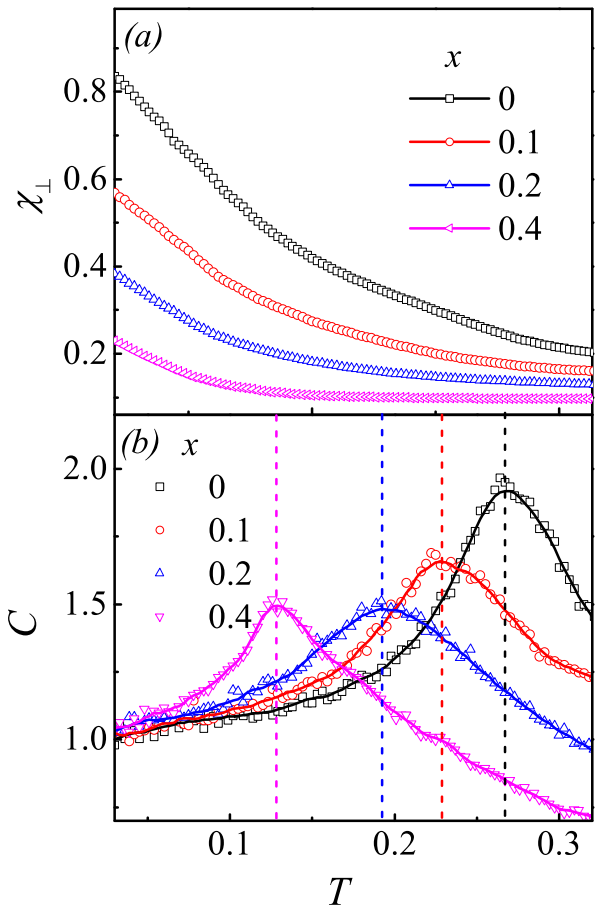

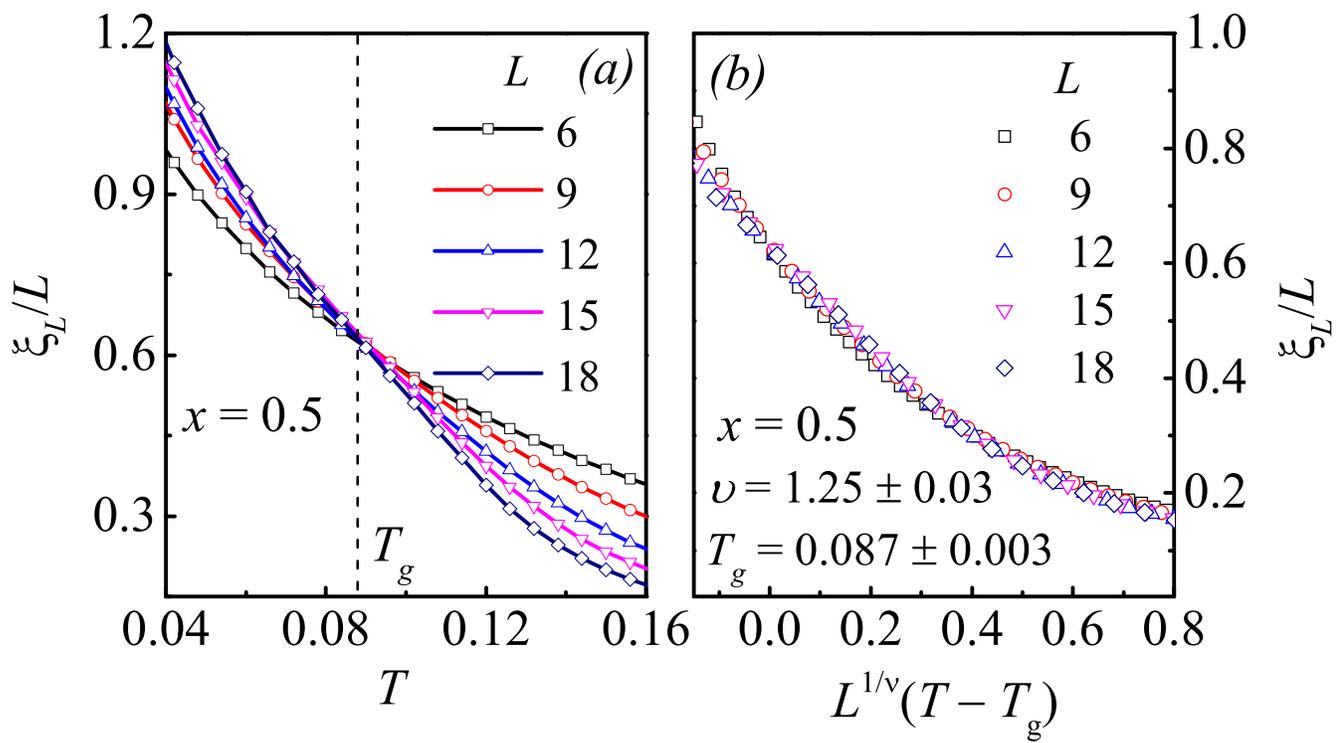

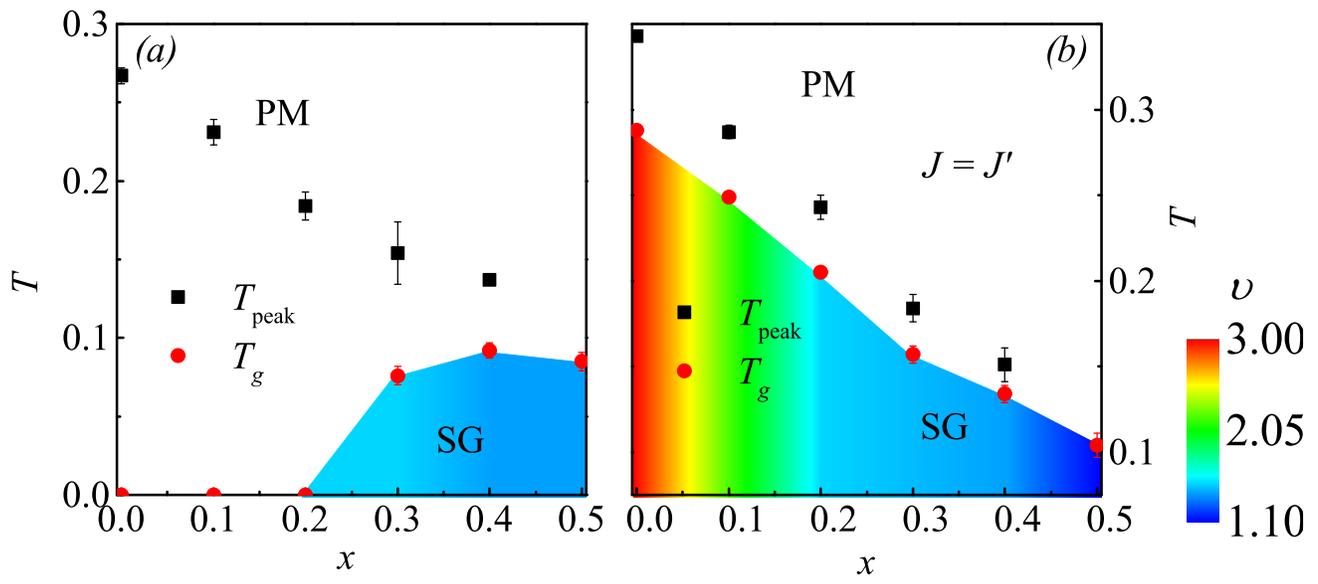

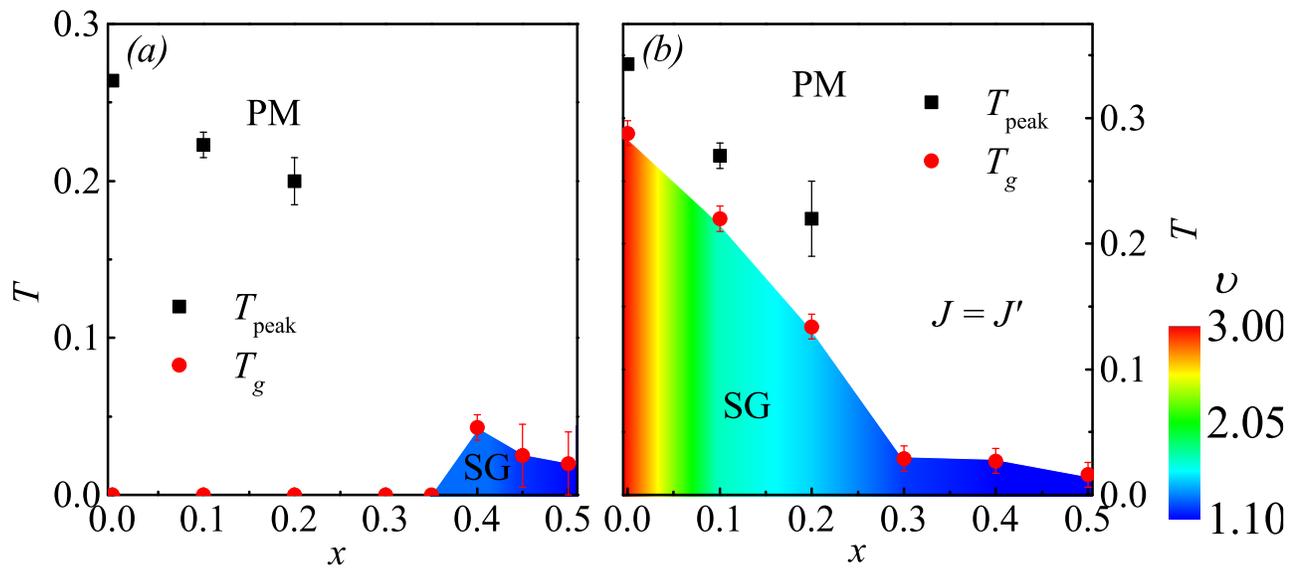

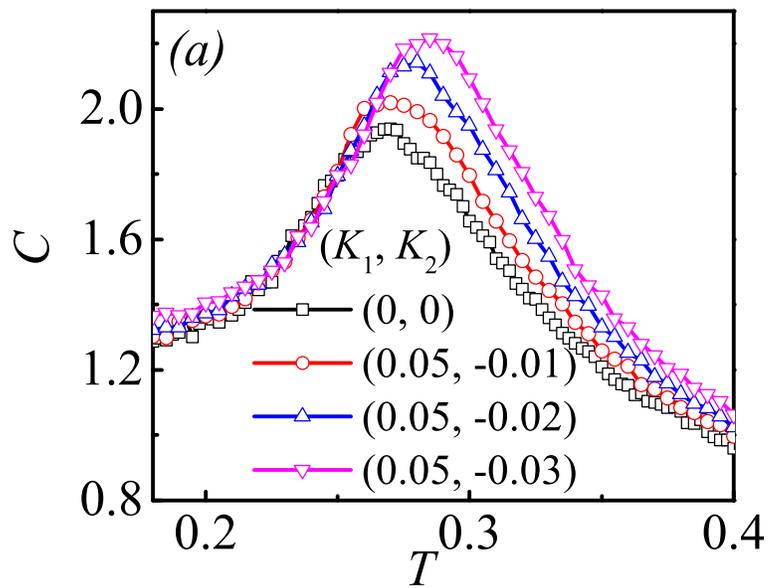
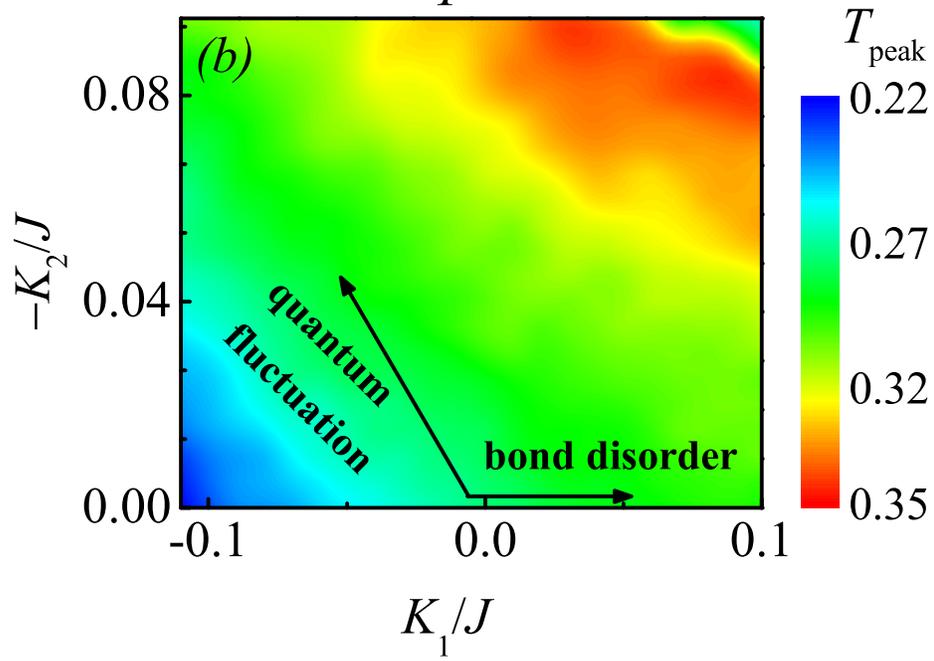

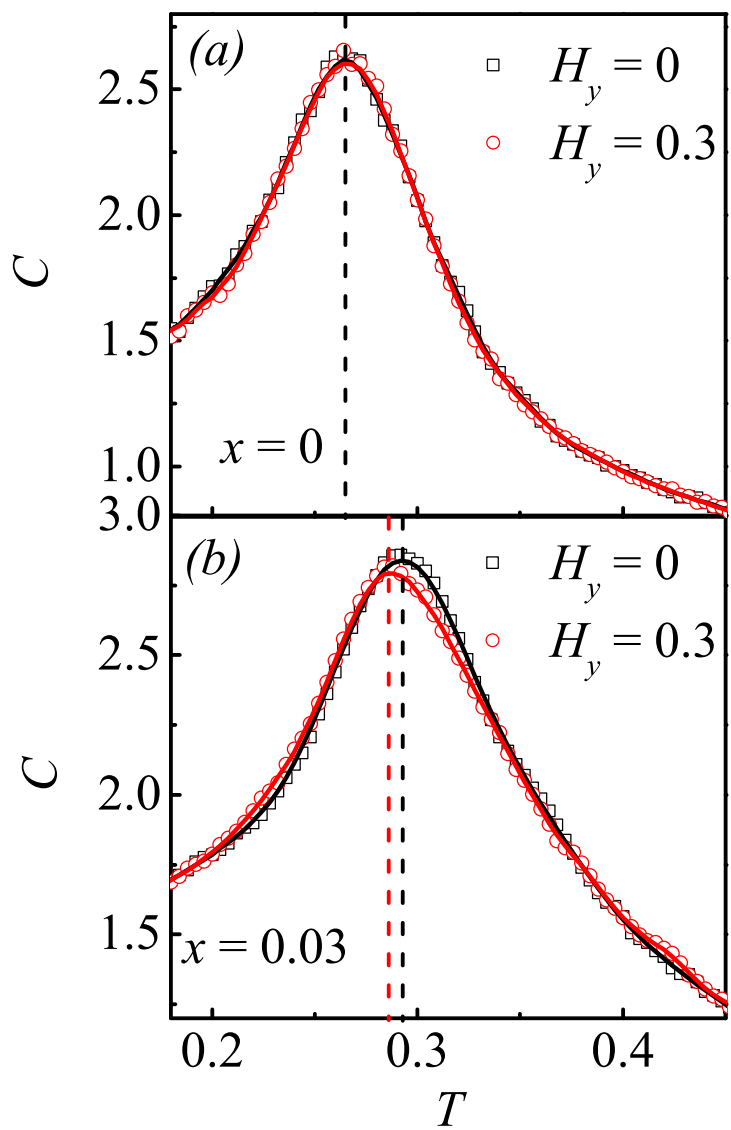